\documentclass[aps,prl,twocolumn,superscriptaddress]{revtex4-2}

\usepackage{amsmath,amssymb}
\usepackage{graphicx}
\usepackage{hyperref}

\begin{document}
	
	\title{The Hidden Second Law of Thermodynamics\\
		behind the Boltzmann-Grad Limit}
	
	\author{Zhaohua Wu}
	\email{zwu@fsu.edu}
	\affiliation{%
		Department of Earth, Ocean, and Atmospheric Science, and \\
		Center for Ocean-Atmospheric Prediction Studies \\ Florida State University \\
		Tallahassee, FL 32306, USA. 
		\vspace{\baselineskip}
	}%
	
	\date{\today}
	
	\begin{abstract}
		We show that the Boltzmann-Grad (BG) limit, $N\varepsilon^{d-1}=\alpha=\text{const}$, is not a neutral scaling condition but already encodes the Second Law of Thermodynamics, and we use this to resolve Loschmidt's paradox. The key to this result is Newton's First Law, which serves as the bridge connecting Newton's Second Law to the Second Law of Thermodynamics. Applying the BG condition locally to two adjacent cells at different temperatures, we decompose their evolution into two stages that make this bridge explicit: a free-streaming period $\delta t_f$, governed by inertia alone (Newton's First Law), during which molecules cross the cell interface ballistically and, having not yet collided, still carry the velocity distribution of their cell of origin; and a subsequent collisional adjustment period $\delta t_a$, governed by Newton's Second Law through the collision dynamics, during which this locally mixed, non-Gaussian population relaxes toward a new Gaussian equilibrium. Using the one-dimensional Gaussian velocity distribution, we show that the isotropy of molecular motion alone produces a non-zero net molecular flux and a non-zero net momentum flux across the interface, both directed from the warmer to the cooler cell -- the microscopic origin of the pressure gradient that drives the system irreversibly toward uniformity. We further give the adjustment period a concrete, computable duration: treating relaxation to Gaussian equilibrium as a sum of $n$ random collisional increments, the Central Limit Theorem fixes the collision count $n^*$ required for convergence, yielding $\delta t_a \sim n^*\,\ell/\bar v$ in terms of the mean free path $\ell$ and mean speed $\bar v$. This two-stage decomposition resolves Loschmidt's paradox by exposing why his reversal argument and Boltzmann's statistical description are not accounts of the same object: Loschmidt's reversal acts on a fixed, identifiable set of molecules, while the temperature field Boltzmann's equation tracks is carried by an ever-changing local population; the step relating the two -- convergence under the Central Limit Theorem -- is itself a coarse-graining that discards microscopic information by construction, and recovering that information carries an unavoidable thermodynamic cost, by Landauer's principle. Temporal irreversibility is not generated during derivation: it is present from the moment the BG framework, and the coarse-graining it requires, are adopted. The spontaneity of Newton's First Law, unfolding without external intervention across the cell interface, is thus the mechanical bridge by which Newton's Second Law and the Second Law of Thermodynamics are shown to be mutually compatible.
	\end{abstract}
	
	\maketitle
	
	\section{INTRODUCTION}
	
	Hilbert's Sixth Problem calls for a rigorous mathematical derivation of the equations of fluid mechanics from the principles of Newtonian mechanics \cite{Hilbert1902}. The central challenge is the derivation of the Boltzmann equation from the $N$-body hard-sphere system, and subsequently the Navier-Stokes equations from the Boltzmann equation.
	
	The BG limit is defined by the scaling condition:
	\begin{equation}
		N\varepsilon^{d-1} = \alpha = \text{const},
		\label{eq:bglimit}
	\end{equation}
	where $N$ is the number of molecules, $\varepsilon$ is the molecular diameter, and $d$ is the spatial dimension. This scaling, introduced by Grad \cite{Grad1949}, defines the dilute gas regime in which the mean collision frequency remains finite as $N\to\infty$ and $\varepsilon\to 0$. It provides the mathematical framework within which a rigorous derivation of the Boltzmann equation from Newtonian hard-sphere dynamics can be attempted.
	
	The first major step toward such a derivation was taken by Lanford \cite{Lanford1975}, who proved that, for random initial conditions distributed according to the BG scaling, the one-particle distribution of the hard-sphere system converges to the solution of the Boltzmann equation -- but only for a very short time, of the order of a fraction of the mean free time between collisions. Extending this result to arbitrarily long times remained an open problem for nearly five decades.
	
	This long-standing gap was recently closed by Deng, Hani, and Ma \cite{DengHaniMa2024}, who provided a rigorous derivation of the Boltzmann equation from hard-sphere dynamics valid for arbitrarily long times, as long as the solution to the Boltzmann equation exists. In a companion paper \cite{DengHaniMa2025}, they subsequently derived the compressible Euler and incompressible Navier-Stokes-Fourier equations from the same microscopic system, completing the resolution of Hilbert's Sixth Problem in the case of a rarefied hard-sphere gas.
	
	Despite this mathematical progress, a fundamental physical question remains unanswered: how does temporal irreversibility -- the arrow of time -- emerge from Newtonian mechanics, which is time-reversible? This is the essence of Loschmidt's paradox \cite{Loschmidt1876}, debated for 150 years.
	
	The standard answer locates irreversibility in the molecular chaos assumption (Stosszahlansatz): the assumption that pre-collision velocities are statistically independent \cite{Boltzmann1872,EhrenfestEhrenfest1959}. This assumption is asymmetric in time -- it holds before collisions, not after -- and thus smuggles in a temporal direction \cite{Cercignani1988,Uffink2007}. However, this raises a deeper question: why should this assumption hold, and where does it come from? We give an explicit, elementary demonstration of what this assumption buys below.
	
	In this Letter, we demonstrate that the answer lies not in the collision operator, but in the BG limit itself. The condition \eqref{eq:bglimit}, applied locally to adjacent cells with different thermodynamic properties, already encodes the Second Law of Thermodynamics through directional molecular exchange. Temporal irreversibility is not generated during derivation -- it is present from the moment the BG framework is established.

	\begin{figure*}
		\centering
		\includegraphics[width=11 cm,height=9cm]{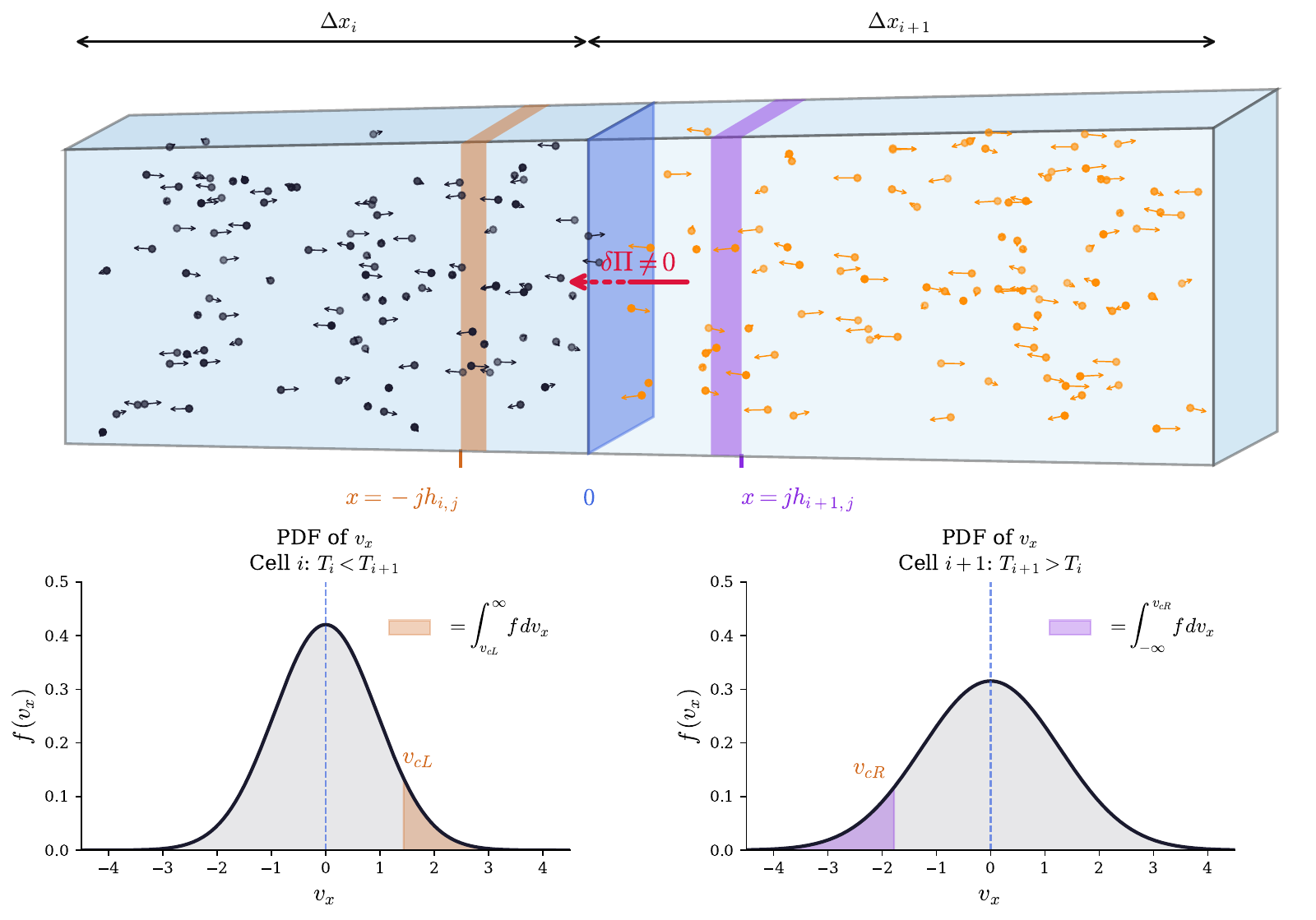}
		\caption{A schematic diagram illustrating the key ideas and results of this paper. In the upper panel, two adjacent cells, each containing $N \to \infty$ molecules in a quasi-equilibrium state, are separated by the blue interface at $x=0$. The probability density function (PDF) below each cell represents the Gaussian distribution of molecular velocities in that cell. The chocolate-colored and the violet blue-colored bands represent the corresponding fractions of each cell after being divided into $M$ sections in the $x$-direction, where $M \to \infty$ but $M/N \to 0$. The chocolate-colored and the violet blue-colored shaded area in each PDF panel shows the molecules whose velocity is sufficiently large to cross the cell interface within $\delta t_{\mathrm{f}}$ from that section. The two shaded areas under PDFs are not equal, meaning the net number of molecules crossing the interface is non-zero. The red arrow in the upper panel represents the net momentum flux $\delta \Pi$, which is non-zero.} 
		\label{schematic}
	\end{figure*}
	
	\section*{TWO ADJACENT CELLS AND THEIR ADJUSTMENTS}
	
	In this section, we consider a macroscopic dilute gas system that has already reached a quasi-equilibrium state at time $t$. We begin from a quasi-equilibrium state purely for expository convenience: the free-streaming and adjustment periods described below in fact repeat indefinitely, with each adjustment period delivering a new quasi-equilibrium state that becomes the starting point of the next free-streaming period, so the choice of $t$ as a starting instant carries no special physical significance. The system is large enough that the following discussion, which focuses on a small interior region of gas over a short time interval $\delta t$, is unaffected by the global boundary conditions, regardless of whether the system is open or closed -- a distinction whose significance will become apparent later. By quasi-equilibrium we mean that the temperature may vary slowly in space, but all other thermodynamic adjustments have equilibrated locally. Since pressure, from a macroscopic perspective, is simply the momentum flux across a unit surface area, its spatial variation at this quasi-equilibrium state -- prior to any further dynamical adjustment -- is negligibly small.
	
	Now we turn our focus to the small interior region. To make the discussion tractable, we adopt a piecewise approach following Deng, Hani, and Ma \cite{DengHaniMa2024}, as illustrated in Fig.~\ref{schematic}, dividing the gas into small imaginary cells, each containing $N\to\infty$ molecules. For simplicity, we consider only a series of small cells in the $x$-direction, assuming that the gas is uniform in the other two spatial directions. The upper panel of Fig.~\ref{schematic} shows two adjacent imaginary cells $i$ and $i+1$ with spatial extents $\Delta x_i$ and $\Delta x_{i+1}$, each independently satisfying the BG condition (1) and containing the same number $N$ of molecules. The cells differ in volume ($\Delta x_i \neq \Delta x_{i+1}$) and hence in temperature and density via the ideal gas law $PV=Nk_BT$, where $k_B$ is the Boltzmann constant. In this setting, since $P_i=P_{i+1}$, and the expansion or contraction of each cell is isotropic in three dimensions, the volume scales as $V_i\propto \Delta x_i^3$, and the ideal gas law $PV=Nk_BT$ requires the temperature to be proportional to $\Delta x_i^3$, or equivalently, $\Delta x_i \propto T_i^{1/3}$.
	
	The boundary between cells is purely mathematical -- it has no physical existence and presents no barrier to molecular motion. Molecules cross freely by the free-streaming mechanism during the time interval $[t,t+\delta t_f]$. Here, $\delta t_f$ is not arbitrary: it represents a short period during which molecules that have crossed the interface have not yet collided with any other molecules in their new cell, and no thermodynamic adjustment within either cell has yet occurred. This free-streaming is not an approximation -- it is a requirement of the BG limit itself, since all collisions are already accounted for in the local probability density function (PDF) of each cell. Under the dilute condition of the BG limit, any additional collision occurring during cross-boundary motion would require modification of the local PDF, violating the BG limit itself. Therefore, free-streaming across the boundary is exact.
	
	We use the one-dimensional Gaussian distribution for the velocity component $v_x$ perpendicular to the boundary:
	\begin{equation}
		f_i(v_x) = \frac{1}{\sqrt{2\pi}\,\sigma_i}\exp\left(-\frac{v_x^2}{2\sigma_i^2}\right),
		\label{eq:gaussian1d}
	\end{equation}
	where $\sigma_i=\sqrt{k_BT_i/m}$ and $m$ is the molecular mass. Note that $\int_{-\infty}^{\infty} f_i\,dv_x = 1$ exactly, as required of a true probability density. We choose the Gaussian rather than the full Maxwell-Boltzmann distribution because cross-boundary exchange is a one-dimensional problem.
	
	\subsection*{Non-Zero Net Molecular Flux}
	
	In this subsection and the next, we prove that the net molecular flux across the interface is directed from the warmer cell to the cooler cell, and so as the momentum flux. To this end, we further divide the $i$-th cell into $M\to\infty$ sections of equal width $h_i=\Delta x_i/M$, with $M/N\to 0$, as illustrated by the chocolate-colored bands in Fig.~1. The condition $M\to\infty$ ensures that the spatial resolution is arbitrarily fine, while $M/N\to 0$ ensures that each section still contains $N/M\to\infty$ molecules, so that the Gaussian velocity distribution \eqref{eq:gaussian1d} remains valid within every section. In this way, we can precisely calculate the number of molecules from each band that cross the interface within the free-streaming period $\delta t_f$.
	
	In time $\delta t_f$, only molecules with $v_x \geq v_{\text{cross},j}$ in the $j$-th section (counted from the interface between the two adjacent cells), where
	\begin{equation}
		v_{\text{cross},j} = \frac{j h_i}{\delta t_f},
		\label{eq:vcross}
	\end{equation}
	can reach and cross the boundary. Therefore, the number crossing from the $j$-th section of the $i$-th cell to the $(i+1)$-th cell is:
	\begin{equation}
		\delta N_{i\to i+1,j} = \frac{N}{M}\int_{v_{\text{cross},j}}^{\infty} \frac{1}{\sqrt{2\pi}\,\sigma_i}\exp\left(-\frac{v_x^2}{2\sigma_i^2}\right)dv_x,
		\label{eq:dN1}
	\end{equation}
	and symmetrically, from the $j$-th section of the $(i+1)$-th cell to the $i$-th cell:
	\begin{equation}
		\delta N_{i+1\to i,j} = \frac{N}{M}\int_{-\infty}^{-v_{\text{cross},j+1}} \frac{1}{\sqrt{2\pi}\,\sigma_{i+1}}\exp\left(-\frac{v_x^2}{2\sigma_{i+1}^2}\right)dv_x,
		\label{eq:dN2}
	\end{equation}
	where $\sigma_i=\sqrt{k_BT_i/m}$ and $\sigma_{i+1}=\sqrt{k_BT_{i+1}/m}$.
	
	Since the two cells are at the same pressure during the free-streaming stage, and the expansion or contraction of each cell is isotropic, the ideal gas law requires their volumes to be proportional to their temperatures. Hence their widths in the $x$-direction satisfy $\Delta x_i/\Delta x_{i+1}=(T_i/T_{i+1})^{1/3}$, and consequently $h_i/h_{i+1}=(T_i/T_{i+1})^{1/3}$.
	
	Introducing the normalized velocities $\tilde v_i = v_x/\sigma_i$ and $\tilde v_{i+1}=v_x/\sigma_{i+1}$, equations \eqref{eq:dN1} and \eqref{eq:dN2} transform to:
	\begin{equation}
		\delta N_{i\to i+1,j} = \frac{N}{M}\int_{\tilde v_{\text{cross}}}^{\infty} \frac{1}{\sqrt{2\pi}}\exp\left(-\frac{\tilde v^2}{2}\right)d\tilde v,
		\label{eq:dN1n}
	\end{equation}
	where $\tilde v_{\text{cross}}=v_{\text{cross},j}/\sigma_i$, and symmetrically:
	\begin{equation}
		\delta N_{i+1\to i,j} = \frac{N}{M}\int_{-\infty}^{-(T_i/T_{i+1})^{1/6}\tilde v_{\text{cross}}} \frac{1}{\sqrt{2\pi}}\exp\left(-\frac{\tilde v^2}{2}\right)d\tilde v.
		\label{eq:dN2n}
	\end{equation}
	
	Since $T_{i+1}>T_i$, we have $(T_i/T_{i+1})^{1/6}<1$, and therefore the upper limit of integration in \eqref{eq:dN2n} satisfies $-(T_i/T_{i+1})^{1/6}\tilde v_{\text{cross}} > -\tilde v_{\text{cross}}$. By the symmetry of the standard Gaussian about zero, this yields $\delta N_{i+1\to i,j} > \delta N_{i\to i+1,j}$ for every section $j$. Summing over all sections:
	\begin{equation}
		\sum_j \delta N_{i+1\to i,j} > \sum_j \delta N_{i\to i+1,j}.
		\label{eq:sumflux}
	\end{equation}
	
	This result establishes that, during the free-streaming period, the number of molecules crossing the boundary from the warmer cell to the cooler cell always exceeds that in the opposite direction. When both cells have identical properties, the two sums are equal and no macroscopic change occurs: this is the equilibrium baseline.
	
	\subsection*{Non-Zero Net Momentum Flux}
	
	Since more molecules cross from the warmer cell to the cooler cell, the net momentum flux is also directed from the warmer cell to the cooler cell. The net momentum transferred across the boundary during $\delta t_f$ is:
	\begin{equation}
		\begin{split}
			\delta\Pi = \frac{mN}{M}\sum_{j=1}^{M}\int_{v_{\text{cross},j}}^{\infty} v_x\, f_i(v_x)\,dv_x \\
			- \frac{mN}{M}\sum_{j=1}^{M}\int_{-\infty}^{-v_{\text{cross},j+1}} v_x\, f_{i+1}(v_x)\,dv_x \\
			\neq 0,
		\end{split}
		\label{eq:momflux}
	\end{equation}
	since $\sigma_{i+1}>\sigma_i$ whenever $T_{i+1}>T_i$, the molecules crossing from the warmer cell carry greater momentum on average than those crossing from the cooler cell. Both the asymmetric molecular flux and the asymmetric momentum per molecule reinforce each other, making $\delta\Pi$ unambiguously directed from the warmer cell to the cooler cell. This net momentum flux is the microscopic origin of the macroscopic pressure gradient between the two cells, which drives the cell interface toward the warmer cell and reduces its volume in the subsequent adjustment stage.

	\subsection*{Timescale Estimates for $\delta t_f$ and $\delta t_a$}
	\label{sec:clt-relaxation}
	
	The classical H-theorem \cite{Boltzmann1872,Maxwell1860} guarantees that, for a closed gas cell, repeated elastic collisions drive the velocity distribution -- regardless of its initial shape -- monotonically toward the unique Gaussian (Maxwellian) equilibrium. It does not, however, supply a timescale for this relaxation. We give here a simpler, physically transparent argument that does: it identifies the number of collisions required for convergence with the number of terms required for a sum of random variables to approach a Gaussian under the Central Limit Theorem (CLT), and then converts that collision count into a physical time using the mean free time between collisions.
	
	Consider two molecules with velocity components $v_1$ and $v_2$ (along the axis of interest, e.g.\ $v_x$), drawn from a common, generally non-Gaussian distribution. In an elastic binary collision -- including general three-dimensional, non-head-on encounters -- the component of the outgoing velocity of a given molecule can be written, to the order relevant here, as a linear combination of the two incoming components,
	\begin{equation}
		V' = a\,v_1 + b\,v_2,
		\label{eq:linearmix}
	\end{equation}
	with coefficients $a,b$ set by the collision (scattering) geometry -- in general, $a$ and $b$ vary from one collision to the next, since they depend on the random impact parameter and scattering angle of that particular encounter. Because successive collisions sample this geometry independently and at random, however, no single encounter's $a,b$ biases the outcome of a long chain of collisions in a fixed direction; it is therefore reasonable, for the purpose of tracking the cumulative statistical effect of many such random encounters, to treat $a$ and $b$ as fixed, representative constants characterizing a typical collision. Following a single molecule through a sequence of $n$ such collisions with partners freshly drawn from the surrounding distribution, its velocity after $n$ collisions is then, to this same order, a linear combination of $n$ independent draws from the original one-molecule distribution:
	\begin{equation}
		V_n \;\sim\; v_1 + v_2 + \cdots + v_n .
		\label{eq:sumform}
	\end{equation}
	Because $N$ is large, the probability that any two of these $n$ collision partners are correlated through a shared earlier encounter is $O(1/N)$ -- the propagation-of-chaos property established rigorously for Kac's random-collision model \cite{Kac1956,McKean1966} -- so the $v_i$ entering the sum may be treated as i.i.d.\ draws from the original distribution to excellent approximation.
	
	Since $v_1,\dots,v_n$ are i.i.d.\ with finite variance $\sigma^2$, the classical CLT guarantees that the distribution of the (normalized) sum $V_n$ converges to a Gaussian as $n$ increases, independently of the shape of the original single-molecule distribution -- reproducing, by an elementary route, the same qualitative conclusion as the H-theorem and as Kac's model \cite{Kac1956}, but tracking the approach to equilibrium directly through the collision count $n$ rather than through the evolution of the full probability density $f(v,t)$.
	
	A quantitative form of the CLT, the Berry--Esseen theorem \cite{Berry1941,Esseen1945}, bounds the distance between the distribution of the normalized sum and the Gaussian by
	\begin{equation}
		D_n \;\lesssim\; \frac{C\,\rho_3}{\sigma^3\sqrt{n}},
		\label{eq:berryesseen}
	\end{equation}
	where $\rho_3$ is the third absolute moment of the original per-molecule distribution and $\sigma$ its standard deviation, so that $\rho_3/\sigma^3$ is a dimensionless measure of how far the initial (post-free-streaming, merged) distribution is from Gaussian. Demanding $D_n \le \eta$ then requires
	\begin{equation}
		n^{*} \;\sim\; \left(\frac{C\,\rho_3}{\sigma^{3}\,\eta}\right)^{2}.
		\label{eq:nstar}
	\end{equation}
	The precise value of $n^{*}$ depends on the degree of non-Gaussianity of the merged distribution (through $\rho_3/\sigma^3$, itself set by the initial temperature contrast $T_{i+1}-T_i$) and on the tolerance $\eta$ demanded of ``Gaussian,'' but its scaling $n^{*}\sim\eta^{-2}$ is universal, and for the modest temperature contrasts relevant to the local, quasi-equilibrium adjustment considered in this Letter, $n^{*}$ is generically $O(1)$--$O(10)$ rather than large.
	
	The mean time between successive collisions of a given molecule, $\tau_c$, is set by the mean free path $\ell$ and the mean molecular speed $\bar v$ \cite{ChapmanCowling1970},
	\begin{equation}
		\tau_c \sim \frac{\ell}{\bar v} .
		\label{eq:tauc}
	\end{equation}
	Since $n^{*}$ collisions per molecule are required for the velocity distribution to relax to within tolerance $\eta$ of the new local Gaussian, the physical adjustment time is
	\begin{equation}
		\delta t_a \;\sim\; n^{*}\,\tau_c \;\sim\; n^{*}(\eta, T_i, T_{i+1})\,\frac{\ell}{\bar v}.
		\label{eq:dta}
	\end{equation}
	This gives the adjustment period introduced in the free-streaming/adjustment cycle a concrete, computable magnitude, expressed entirely in terms of measurable quantities -- mean free path, mean speed, and the initial temperature contrast between the merging cells -- rather than left as a qualitative stage.
	
	$\delta t_f$, by contrast, was defined as the interval during which molecules that have crossed the cell interface have not yet undergone their first collision in the new cell. This is, by construction, of order a single mean free time,
	\begin{equation}
		\delta t_f \;\sim\; \tau_c,
		\label{eq:dtf}
	\end{equation}
	in contrast to $\delta t_a$, which requires the accumulation of $n^{*}\!\gtrsim\!1$ such collisions to complete the relaxation to the new Gaussian. Because $n^{*}\ge 1$, the adjustment period is generically the slower of the two stages -- consistent with its role as the stage that actually establishes the new, merged local temperature -- while both are governed by the same microscopic timescale $\tau_c$, so that the total step $\delta t = \delta t_f + \delta t_a \sim (1+n^{*})\,\tau_c$ remains, self-consistently, of order a few mean collision times rather than some separate, unrelated macroscopic scale.
	
	It is worth noting that the resulting picture -- a free-streaming stage followed by a discrete count of $n^{*}$ collision events, iterated cell by cell -- bears a resemblance to a Riemann sum approximating an integral, with $\tau_c$ playing the role of the discretization step and $\delta t_a$ accumulating in units of it. We do not claim this correspondence rigorously; unlike a Riemann sum, collisions here are not evenly spaced, discrete events imposed on an otherwise continuous process but rather occur continuously and asynchronously across the ensemble of molecules at all times. The discrete count $n^{*}$ should therefore be understood as a mean-field bookkeeping device for estimating a timescale, not as a literal discretization of the underlying continuous-time collision process.
	
	This argument idealizes the true collision kinematics -- the linear-mixing ansatz $V'=av_1+bv_2$ [Eq.~\eqref{eq:linearmix}] suppresses the dependence of the exact three-dimensional elastic-collision outgoing velocity on the relative speed and the random scattering direction -- and does not aim to reproduce the rigorous convergence-rate results available from Kac's master-equation formalism and its spectral-gap analysis \cite{Kac1956,McKean1966,Villani2003}. Its value lies instead in supplying a transparent, physically motivated, and directly numerically verifiable (e.g.\ by Monte Carlo simulation of repeated random collision pairings, checked against a Kolmogorov--Smirnov test) order-of-magnitude estimate for $\delta t_a$, expressed in exactly the microscopic quantities ($\ell$, $\bar v$, and the initial non-Gaussianity) that the piecewise cell-iteration scheme of this Letter requires. As a physicist's estimate -- an order-of-magnitude timescale grounded in measurable microscopic quantities, rather than a sharp mathematical bound -- this level of rigor is fit for the purpose it serves here.

	\section*{THE HIDDEN SECOND LAW OF THERMODYNAMICS}
	
	Both the net molecular flux and the net momentum flux are directed from the warmer cell to the cooler cell. These imbalances drive the system irreversibly toward uniformity. In principle, we can calculate the macroscopic properties of each cell over a complete episode of duration $\delta t$, consisting of a free-streaming period $\delta t_f$ and a subsequent adjustment period $\delta t_a$, following these steps:
	
	\begin{enumerate}
		\item Compute the net molecular flux and net momentum flux at the end of the free-streaming period $\delta t_f$.
		\item Update the number of molecules, pressure, volume, and temperature of each cell via $PV=Nk_BT$.
		\item Compute the new PDFs from the updated thermodynamic states at the end of the adjustment period $\delta t_a$.
		\item Reapply the BG condition (1) to obtain a new set of cells and repeat.
	\end{enumerate}
	
	Each iteration brings the two cells closer to uniformity. The process is strictly irreversible. This is the Second Law of Thermodynamics -- not postulated, not derived from molecular chaos, and not dependent on special initial conditions, but following necessarily from the local BG limit, the isotropic nature of molecular motion, free-streaming, and the iterative update of the PDFs.
	
	During the free-streaming period, molecules crossing from the higher-temperature cell to the lower-temperature cell carry, on average, greater momentum and energy than those crossing in the opposite direction -- a direct consequence of the higher molecular speeds in the hotter cell. During the subsequent adjustment period $\delta t_a$, collisions within each cell redistribute this transferred momentum and energy among all molecules, raising the macroscopic temperature of the cooler cell and lowering that of the hotter cell.
	
	The physical essence of the result can be stated without counting molecules. During the free-streaming period, molecules from the cooler cell -- carrying smaller average momentum -- enter the warmer cell, while molecules from the warmer cell -- carrying larger average momentum -- enter the cooler cell. This mutual exchange inevitably pulls the average molecular speeds of the two cells toward each other. The precise number of molecules exchanged determines only the rate at which the temperature difference diminishes, not the direction: heat always flows from the warmer cell to the cooler cell. This is Clausius's Second Law \cite{Clausius1850,Clausius1854}, and it follows from nothing more than the free inertial motion of molecules across an imaginary boundary.
	
	Our framework reveals the microscopic origin of Clausius's observation. This exchange is asymmetric in momentum and energy. Heat flows from hot to cold not because nature has any mysterious preference, but because molecules in the absence of external forces continue their inertial motion: Newton's First Law, and nothing more.
	
	\begin{figure*}
		\centering
		\includegraphics[width=10 cm,height=10cm]{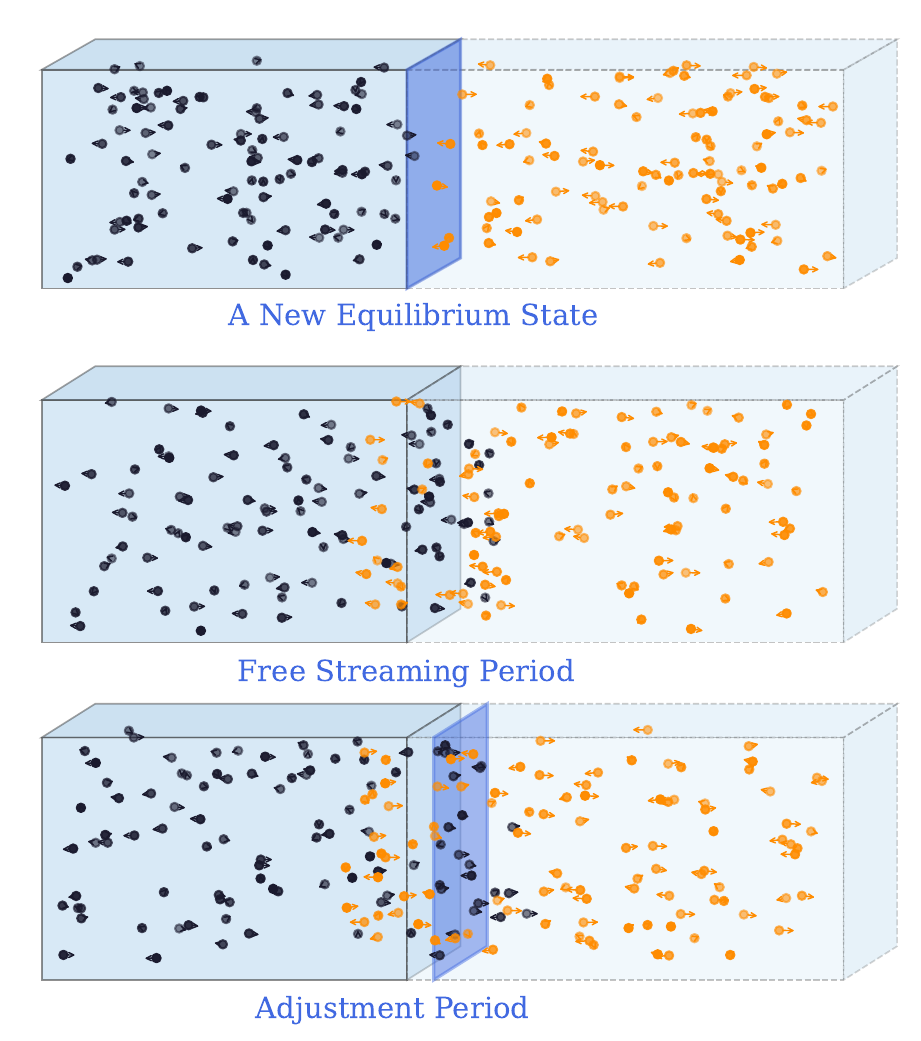}
		\caption{Two adjoining cells, each containing $N$ molecules, separated by an imaginary wall whose position shifts so as to always leave $N$ molecules on each side -- consistent with the flux calculation. Black/orange dots denote molecules originally native to the left/right cell. \emph{Free Streaming} (middle): molecules that have just crossed the wall retain their cell of origin's velocity distribution. \emph{Adjustment} (bottom): collisions between native and crossed molecules relax the locally mixed, non-Gaussian distribution toward a new Gaussian (\S\ref{sec:clt-relaxation}). \emph{New Equilibrium} (top): the resulting population is fully mixed and Gaussian about an intermediate temperature. The three panels show that the local molecular population is not fixed in time.} 
		\label{schematic2}
	\end{figure*}
	
	\section*{RESOLUTION OF LOSCHMIDT'S PARADOX}
	\label{sec:loschmidt-mismatch}
	
	Figure~\ref{schematic2} makes visible why Loschmidt's reversal argument and Boltzmann's statistical description, despite appearing to address the same physical situation, are in fact not competing accounts of the same object -- and why Loschmidt's paradox therefore supplies no valid logical basis for objecting to the Boltzmann distribution.
	
	Macroscopically, temperature is defined as a \emph{field}: at each point in space it characterizes the statistical distribution of the velocities of whichever molecules currently occupy that point. It carries no record of, and does not track, any particular set of molecules that occupied that point earlier. As the middle panel of Fig.~\ref{schematic2} (``Free Streaming Period'') shows, once molecules begin crossing a cell boundary, the population occupying a given local region is, after a free-streaming interval, already a \emph{different} set of molecules from the one that occupied it before: some original occupants have left, and molecules from the neighboring, differently-tempered region have entered. The relaxation described by the H-theorem -- and by the CLT argument of \S\ref{sec:clt-relaxation} above -- therefore acts on this new, mixed population, not on the original molecules that Loschmidt's reversal operation would act on. Reversing only the velocities of the original set of molecules is consequently far from sufficient to reverse the macroscopic evolution once mixing has occurred. If a genuine reversal is to be achieved after sufficient time has elapsed, it is not enough to reverse a local subset of molecules: it requires reversing the velocities of every molecule of the entire isolated system, simultaneously. But temperature, by its very definition as a field, is local, and its down-gradient propagation reflects only the local, ever-changing subset of molecules currently present -- never the fixed, global population that a genuine Loschmidt reversal would have to act on.
	
	This can be sharpened into a direct, elementary demonstration of what the molecular chaos assumption buys. Consider a single instant during a cell's quasi-equilibrium period, and let $S$ denote the specific phase-space state of its molecules -- each with a definite position and velocity -- and let $S^{*}$ denote its time-symmetric partner, obtained by reversing every velocity while leaving every position unchanged. Because the local velocity distribution $f_i(v_x)$ [Eq.~\eqref{eq:gaussian1d}] is Gaussian and therefore an even function of $v_x$, at every position within a section the number of molecules with velocity $+v_x$ equals, by isotropy, the number with velocity $-v_x$. Repeating the flux calculation above starting from $S^{*}$ in place of $S$ therefore yields exactly the same crossing count and the same net momentum flux, both still directed from the warmer cell to the cooler cell -- with a different set of molecules doing the crossing, but the same macroscopic direction and magnitude. This equality is not merely an ensemble average: it follows from the isotropy of the local Gaussian distribution alone, for an arbitrary specific microstate $S$.
	
	This does not conflict with Loschmidt's reversibility objection, discussed above; it is instructive precisely because of why it holds. $S^{*}$ reproduces the same macroscopic flux because $S$, drawn from a quasi-equilibrium, molecular-chaos-obeying ensemble, carries no built-in correlation between a molecule's position and its velocity: reversing velocities while leaving positions fixed simply produces another molecular-chaos-obeying Gaussian state, statistically indistinguishable from the first. A genuine Loschmidt reversal of an already-mixed, history-correlated population -- the scenario addressed via Poincar\'e recurrence below -- is a different operation: it reverses a state whose position-velocity correlations were built up by the preceding dynamics, and it is precisely those correlations, absent here, that would let the reversed trajectory retrace the unmixing. The elementary calculation above thus gives concrete, computable content to the traditional observation that molecular chaos, not microscopic irreversibility, is where the arrow of time enters: as long as the local ensemble remains uncorrelated in this sense, time-reversing it changes only which molecules cross the interface, never the direction in which they do.
	
	This physical mismatch has a precise mathematical counterpart in the CLT machinery, and it is worth being explicit that the two are not merely analogous but identical in substance, in the coarse-graining sense long employed in statistical mechanics \cite{Tolman1938,Jaynes1957}. Two layers of coarse-graining are involved. First, the passage from the individual velocities $v_1,\dots,v_n$ of a specific, identifiable set of colliding molecules to their sum $V_n=v_1+\cdots+v_n$ is itself a coarse-graining: it is a marginalization that discards the joint microscopic information -- which particular molecule carried which velocity -- and retains only the aggregate. This is the same operation, in kind, as replacing a system's microstate with a macroscopic aggregate such as its total energy or momentum. Second, the limiting distribution that the CLT assigns to (the normalized) $V_n$ discards further information as $n\to\infty$: writing the density of the standardized sum via its Edgeworth expansion,
	\begin{equation}
		\begin{split}
			f_{V_n}(x) \;\approx\; \phi(x)\bigg[\,1 \;+\; \frac{\gamma_1}{6\sqrt{n}}\,He_3(x) \\
			\;+\; \frac{\gamma_2}{24\,n}\,He_4(x) \;+\; O(n^{-3/2})\,\bigg],
		\end{split}
		\label{eq:edgeworth}
	\end{equation}
	where $\phi$ is the standard normal density, $He_k$ the probabilists' Hermite polynomials, and $\gamma_1,\gamma_2$ the skewness and excess kurtosis of the original per-molecule distribution, one sees explicitly that the shape information distinguishing one original distribution from another -- carried by $\gamma_1,\gamma_2$, and by all higher cumulants beyond them -- is suppressed by successive powers of $n^{-1/2}$ and vanishes identically in the $n\to\infty$ limit, consistent with the Berry--Esseen bound of Eq.~\eqref{eq:berryesseen}. Only the first two moments, mean and variance, survive into the limiting Gaussian. Many distinct initial (non-Gaussian) populations, mixed and relaxed through collisions, therefore converge to the exact same Gaussian; the CLT is many-to-one, and the information distinguishing one such population from another is discarded in the limit, not merely hidden. The CLT is, in this precise sense, the mathematical expression of coarse-graining itself -- which is exactly why demanding that it be run backward, as Loschmidt's reversal implicitly does, is not a request to reverse a physical process but a request to invert a map that discards information by construction.
	
	Loschmidt's reversal objection is, taken at face value, entirely correct: since the dynamics is time-reversible, simultaneously reversing every molecule's velocity in the isolated system does return it, along the exact retraced trajectory, to its original, unmixed configuration -- and Poincar\'e's recurrence theorem \cite{Poincare1890} guarantees that, given sufficient time, the system's own forward dynamics will on its own eventually bring it arbitrarily close to that same configuration again, with no external reversal required at all. Boltzmann's own reply to this objection, and to Zermelo's closely related recurrence objection \cite{Zermelo1896}, was itself probabilistic, and it remains correct: the reversed (or recurred) configuration corresponds to a single, essentially measure-zero point in the space of microstates compatible with the system's macroscopic constraints, against the overwhelming statistical weight of the mixed, Gaussian configuration \cite{Lebowitz1993}. The CLT argument above gives this weighting a concrete mechanism, rather than leaving it as a counting exercise: the map from individual molecular velocities to their locally averaged, mixed distribution is many-to-one, so an astronomically large set of distinct microstates is compatible with the mixed Gaussian macrostate, while the exactly reversed or recurred configuration is one specific, atypical point in that same space. Should the system's own dynamics ever carry it through that point, it does so only instantaneously, immediately resuming its overwhelmingly more probable evolution back toward the mixed state. Boltzmann's probabilistic answer and the coarse-graining argument above are thus the same statement in two languages: one counts microstates, the other identifies the many-to-one map responsible for the counting -- and both confirm that Loschmidt's reversal and Boltzmann's H-theorem were never in genuine tension to begin with. This has a sharp corollary: since the system's own dynamics carries it through the reversed configuration only for a fleeting instant before entropy resumes increasing, sustaining any departure from this overwhelmingly probable evolution -- rather than a single momentary crossing -- requires continuous intervention from outside the isolated system, not a one-time reversal.
	
	This informational mismatch has a further, thermodynamic dimension, made precise by Szilard's and Landauer's analysis of Maxwell's Demon. The information discarded by the coarse-graining above is not merely a conceptual simplification adopted by the observer: recovering it, and later disposing of it, carries an unavoidable thermodynamic cost. By Landauer's principle \cite{Landauer1961}, erasing one bit of information dissipates at least $k_BT\ln 2$ of heat; any agent that acquires the microscopic information needed to genuinely reverse the mixed population of the preceding paragraphs -- rather than merely flipping velocities without regard to which molecules they belong to -- must eventually erase that information to complete any thermodynamic cycle, generating at least as much entropy as the reversal would appear to remove. This is precisely the mechanism by which Landauer, and later Bennett \cite{Bennett1982}, building on Szilard's original analysis of his engine \cite{Szilard1929}, resolved the Maxwell's Demon paradox; it applies here with the same force. Even granting that the underlying dynamics is exactly time-reversible -- that nothing in Newton's laws themselves forbids the reversal -- actually acquiring and using the microscopic information that coarse-graining discarded is not thermodynamically free. The reversal is not merely difficult to arrange in practice; using the information it requires necessarily produces entropy elsewhere, by the same principle that saves the Second Law from Maxwell's Demon.
	
	Taken together, these three layers of argument -- dynamical, statistical, and thermodynamic -- show that Loschmidt's reversal and Boltzmann's H-theorem are not two accounts of the same process read differently. The dynamics is exactly reversible, as Loschmidt correctly observed; but the statistical description Boltzmann's equation tracks is built on a coarse-graining -- the CLT step -- that discards information by construction, and any attempt to physically undo that discarding carries an irreducible thermodynamic cost. Because the two arguments do not share a common object, Loschmidt's paradox is not resolved by reconciling reversibility with irreversibility -- it was never actually addressed to the target it is commonly read as refuting.
	
	\section*{CONCLUSION}
	
	The central finding of this work is not merely the resolution of Loschmidt's paradox, but the demonstration that Newton's First Law, Newton's Second Law, and the Second Law of Thermodynamics are mutually compatible -- with Newton's First Law serving as the bridge between the other two, and the Central Limit Theorem supplying that bridge's quantitative timescale. The spontaneity of inertial motion, acting across the imaginary boundary between cells of differing temperature, produces an asymmetric momentum exchange during a free-streaming period $\delta t_f$; the subsequent adjustment period $\delta t_a$, during which collisions relax the resulting mixed, non-Gaussian population to a new local Gaussian, completes the cycle. This asymmetry is the Second Law of Thermodynamics -- not in conflict with Newtonian mechanics, but an expression of it at the statistical level.
	
	The Boltzmann-Grad limit $N\varepsilon^{d-1}=\alpha$ is not a neutral mathematical scaling condition. Applied locally to adjacent cells with differing thermodynamic properties, it necessarily encodes:
	
	\begin{itemize}
		\item Non-zero net molecular flux due to isotropic expansion: since $\Delta x_i \propto T_i^{1/3}$, we have $h_{i+1}/h_i = (T_{i+1}/T_i)^{1/3} \neq \sqrt{T_{i+1}/T_i}$, so that more molecules cross from the hotter cell to the cooler cell;
		\item Non-zero net momentum flux (the microscopic pressure gradient, $\delta\Pi \neq 0$);
		\item A concrete, computable adjustment timescale $\delta t_a \sim n^*\,\ell/\bar v$, obtained by treating the relaxation to Gaussian equilibrium as governed by the Central Limit Theorem, rather than left as a qualitative stage;
		\item Irreversible evolution toward thermodynamic uniformity (the Second Law of Thermodynamics).
	\end{itemize}
	
	The Boltzmann equation is Newton's Second Law in phase space, with the Second Law of Thermodynamics built into its structure through the BG limit. The collision operator $Q(f,f)$ is the macroscopic force of molecular exchange. The relationship between the Boltzmann and Navier-Stokes equations is scale transformation, not derivation. Loschmidt's paradox dissolves: temporal irreversibility was never absent from the framework.
	
	At the microscopic level, for a dilute gas, Clausius's Second Law of Thermodynamics -- that heat flows spontaneously from hot to cold -- is simply a manifestation of the spontaneity of Newton's First Law. Molecules crossing the interface from the hotter cell carry greater momentum and energy; those crossing from the cooler cell carry less. Moreover, more molecules cross from the hotter cell than from the cooler cell. The exchange is asymmetric in both number and momentum.
	
	This irreversibility can be seen directly from the structure of the Navier-Stokes equations themselves. Since the Navier-Stokes equations are derived from the Boltzmann equation \cite{DengHaniMa2025}, which in turn inherits the Second Law of Thermodynamics already encoded in the BG limit, the irreversibility is not an artifact of the fluid equations but a property carried through from the microscopic framework. The advection term $-u\,\partial u/\partial x$ always acts down the gradient, transferring momentum from regions of higher velocity to regions of lower velocity. One might object that replacing $\delta t$ with $-\delta t$ also reverses the sign of this term, apparently restoring the original tendency. However, after forward integration, the velocity gradient has already been reduced. The reversed advection term therefore acts on a smaller gradient, producing insufficient change to recover the initial state: the asymmetry lies not in the sign of the tendency, but in its magnitude. Requiring up-gradient transfer is thus a physical impossibility already forbidden by the equation's own structure. Furthermore, since the advection term always acts to smooth velocity gradients rather than amplify them, the formation of singularities -- which would require infinite amplification of gradients in finite time -- appears to be resisted by the same mechanism. This suggests that the existence of smooth solutions to the Navier-Stokes equations, one of the Clay Millennium Problems, may follow from the Second Law of Thermodynamics already encoded in the equations' structure.
	
	The true source of the paradox does not lie in a conflict between reversible dynamics and irreversible thermodynamics: the underlying dynamics is exactly time-reversible, as Loschmidt correctly observed. It lies instead in three compounding mismatches, made explicit above. Dynamically, Loschmidt's reversal requires the simultaneous, external reversal of a fixed, identifiable set of molecules, whereas inertial motion is spontaneous and admits no internal mechanism for such a reversal to occur on its own. Statistically, the temperature field Boltzmann's equation tracks is carried by an ever-changing local population, related to any fixed initial population only through the Central Limit Theorem -- itself a coarse-graining that discards microscopic information by construction, not a process that can be run backward. Thermodynamically, recovering the information that coarse-graining discards, as required to genuinely undo it, carries an unavoidable entropic cost, by the same Landauer--Szilard--Bennett principle that resolves the Maxwell's Demon paradox. Loschmidt's argument addresses a system that the Boltzmann equation was never describing.

	\bigskip
	
	\begin{acknowledgments}
		\textit{Acknowledgments}---The author benefited from extensive discussions with Claude (Anthropic) in clarifying the presentation of the work. The conception, physical interpretation, mathematical development, and scientific conclusions are entirely the author's own.
		
	\end{acknowledgments}
	

	\end{document}